\documentclass{article}
\usepackage{times, mathptmx, amssymb,graphicx,amsmath, axodraw, epsfig, psfrag, color}

\def\mt1{\ensuremath{\tilde{m}_1}}
\def\MWR{\ensuremath{M_{W_R}}}
\def\be{\begin{equation}}
\def\ee{\end{equation}}

\definecolor{rouge}{rgb}{1,0,0}

\newcommand{\preprintno}[1]{\vspace{-2cm}{\normalsize\begin{flushright}#1\end{flushright}}\vspace{1cm}}

\newenvironment{narrow}[2]{%
\begin{list}{}{
\setlength{\leftmargin}{#1}%
\setlength{\rightmargin}{#2}%
\setlength{\listparindent}{\parindent}%
\setlength{\itemindent}{\parindent}%
\setlength{\parsep}{\parskip}}%
\item[]}{\end{list}}%

\makeatletter
\newcommand\figcaption{\def\@captype{figure}\caption}
\newcommand\tabcaption{\def\@captype{table}\caption}
\makeatother

\author{
N. Cosme\thanks{ncosme@ulb.ac.be},
\\
\textit{\normalsize{Service de Physique Th\'eorique, CP225}}\\
\textit{\normalsize{Universit\'e Libre de Bruxelles}},\\
\textit{\normalsize{Bld du Triomphe, 1050 Brussels, Belgium.}}}

\title{Leptogenesis, neutrino masses and gauge unification
\preprintno{{\bf ULB-TH/04-07.}}
 }

\begin{document}
\date{}
\maketitle

\begin{abstract}
Leptogenesis is considered in its natural context where Majorana neutrinos fit in a gauge unification scheme
and therefore couple to some extra gauge bosons. The masses of some of these gauge bosons are expected to be similar to those of the heavy Majorana particles, and this can have important consequences for leptogenesis.
In fact, the effect can go both ways. Stricter bounds are obtained on one hand due to the dilution of the CP-violating effect by new decay and scattering channels, while, in a re-heating scheme, the presence of gauge 
couplings facilitates the re-population of the Majorana states. 
The latter effect allows in particular for smaller Dirac couplings.
\end{abstract}


\section{Introduction}

Thermal leptogenesis \cite{Fukugita:1986hr} takes place through CP-violating decay of heavy neutrinos 
in the evolution of the early universe. It eventually results in the
baryon/antibaryon imbalance through partial conversion of leptons to baryons 
around the electroweak scale. From Sakharov's conditions \cite{Sakharov:dj},
successful leptogenesis depends on both the amount of CP-violation and the
importance of the departure from equilibrium.

In the simple model, i.e. the Standard Model (SM) together with three singlet heavy Majorana neutrinos, the description of these phenomena can be accurately described by few
parameters: light neutrino masses
\footnote{that is more precisely, the quantity $\mt1 =\left( \lambda_\nu
\lambda_\nu^\dagger\right)_{11} v^2/M_1$ for $\lambda_\nu$ the left-handed
neutrino Yukawa couplings matrix and $M_1$ the lightest Majorana mass.}  and the
lightest Majorana mass  \cite{Barbieri:1999ma}, \cite{Plum2003}, \cite{Giudice} (assuming an important hierarchy between Majorana states).

The inclusion of right-handed Majorana neutrinos is however quite difficult to justify outside
some Grand Unification context (the most obvious candidate being $SO(10)$). Obviously, the unification is accompanied by gauge bosons coupled to
Majorana neutrinos. Irrespective of the detailed breaking scheme, we can expect that the masses of some 
gauge bosons will be linked to the breaking mechanism associated to the
Majorana mass. As a consequence, a full description of leptogenesis should at the very least consider the part 
of the gauge sector linked to right-handed neutrinos, that is particles as $W_R^\pm$ and a $Z^\prime$.

While we insist that the actual breaking scheme does not need to include an explicit passage 
through the stage $SU(2)_R \times U(1)$, it is logical to include at least the above-mentioned particles in the analysis. 

The inclusion of such particles can have non negligible impact 
on the analysis. The most straightforward is probably an extra dilution of the generated CP violation \cite{ling-frere}, 
and hence ultimately baryon number due to new decay and scattering channels. This will be 
shown in some detail below. On the other hand, the new gauge couplings will also be seen to enhance leptogenesis 
in some part of the parameter space, namely when the Yukawa couplings (Dirac masses) are small. In this case indeed, 
the presence of the larger gauge interactions favor the production of the heavy Majorana particles in a re-heating 
context. 

As a result of both effects, we will thus find that the parameter area is both shrunk on one side, and extended to 
low Yukawa coupling values.

To be definite, and despite the above remark that the present considerations apply very generally in unified theories, 
whether the LR-symmetric stage is actually realized or not as an intermediary step of the symmetry breaking, we will use
 the simplest consistent structure, namely 
$SU(2)_L \times SU(2)_R \times U(1)_{B-L}$ (note that the coupling constants of the two $SU(2)$ groups need not be equal,
 but this translates only below in a re-definition of the gauge  mass scale).

Also for definiteness, we assume that the $SU(2)_R \times U(1)_{B-L}$ is broken by a scalar triplet of $SU(2)_R$; 
we have also assumed that the mass of the surviving scalars are heavy enough for their effect to be neglected here 
(this scheme is opposite to the one considered in \cite{vero}).
This scalar $VEV$ as usual provides masses to $W_R^\pm$ and $Z^\prime$ gauge bosons and a 
Majorana mass for right-handed neutrinos $N$. 
The high Majorana mass leads to the "see-saw" mechanism \cite{SeeSaw} for the 
left-handed neutrino mass through the Yukawa couplings:
\be \bar{l}_L \phi \lambda_l \; e_R + \bar{l}_L \tilde{\phi} \lambda_\nu \; N +\frac{1}{2} \bar{N^c} M N +h.c.,\ee
where $\phi$ is the SM scalar doublet, and $l_L=\left(\begin{array}{cc} \nu_L & e_L \end{array}\right)^T$.

As, announced, the expected effects include:
\begin{enumerate}
\item The gauge dilution of the CP-asymmetry due to a CP-invariant decay rate through the gauge sector 
\cite{ling-frere}, this will be dealt with in section \ref{CP}.
\item A lower Majorana decoupling due to new diffusion reactions 
beside the one usually considered in the minimal case cf. Fig \ref{phi-diffusion2}, \ref{phi-diffusion1}.
These additional diffusions keep heavy neutrinos at equilibrium in a wider parameter space, namely for
small Majorana masses as seen below in section \ref{Matter}.
\item In a re-heating scenario, the probability to produce the heavy neutrino can be considerably enhanced by their 
gauge coupling with respect to the minimal case. This point is also discussed in section \ref{Matter}.
\end{enumerate}

\section{CP-Asymmetry}\label{CP}

The lepton asymmetry originates from the leptonic number violating and CP violating decay of Majorana neutrinos
($N_1 \to l_L \; \phi$ +  $N_1 \to \bar{l}_L \; \phi^\dagger$).
The CP asymmetry in the decay is generated at the one loop level thanks to the complex nature of the neutrino
mass matrix. In the minimal model
of leptogenesis, this CP asymmetry for the lightest Majorana can be written as:
\be \epsilon_1= \frac{\Gamma(N_1 \to l \; \phi)-\Gamma(N_1 \to \bar{l} \; \phi^\dagger)}
{\Gamma(N_1 \to l \; \phi)+\Gamma(N_1 \to \bar{l} \; \phi^\dagger)}=
-\frac{3}{16 \pi} \frac{1}{\left(\lambda_\nu \lambda_\nu^\dagger\right)_{11}} \sum_{i=2,3} \frac{M_1}{M_i}
Im\left[\left(\lambda_\nu \lambda_\nu^\dagger\right)^2_{1i} \right],\ee
in the basis of diagonal Majorana mass matrix, with $M_1 \ll M_2 \ll M_3$.

Even though $\epsilon_1$ depends on the flavor structure of the Yukawa couplings, and hence of the different
unknown mass parameters, the Davidson-Ibarra  bound \cite{DI} gives:
\be |\epsilon_1| < \frac{3}{16 \pi} \frac{M_1}{v^2} \left( m_3 -m_1 \right).\ee
Moreover, an improvement has been recently proposed in \cite{Hambye}, assuming the neutrino oscillation parameters 
$\Delta m^2_{solar} << \Delta m^2_{atm}$, that is:
\be |\epsilon_1| < \frac{3}{16 \pi} \frac{M_1}{v^2} \left( m_3 -m_1 \right) \frac{1}{2}
\sqrt{1-\left[\frac{(1-a) \mt1}{(m_3-m_1)}\right]^2} \sqrt{(1+a)^2 - \left[\frac{(m_3+m_1)}{\mt1}\right]^2}, \label{eH}\ee
with $a= 2 \mathcal{R}e\left[\frac{m_1 m_3}{\mt1^2}\right]^{1/3} \left[-1-i \sqrt{\frac{(m_1^2+m_3^2+\mt1^2)^3}
{27 m_1^2 m_3^2 \mt1^2}-1}\right]^{1/3}$.

This bound provides an upper estimate of the CP asymmetry based on experimental constraints, and 
irrespective of the specific mass pattern assumed in solar and atmospheric neutrino oscillations.

Let us turn to the decay channels for the
Majorana neutrinos mediated by the extended gauge structure. Using the familiar notation $N_1$ for the lightest Majorana state, of mass $M_1$, two cases can be distinguished.
When $M_1 > M_{W_R}$, i.e. when the 
$W_R$ is lighter than all Majorana particles, the additional channels will mostly occur in a two body decay into an on-shell $W_R$, and will be found to dilute excessively the induced CP asymmetry.
For $M_1 < M_{W_R}$ instead,  
the three-body decay channels will dominate.

These mostly CP conserving channels will result in a diluted CP asymmetry with respect to the minimal model, which we parametrise by $X$:
\be \epsilon_1^{tot}= \frac{\epsilon_1}{1+X},\ee
where the total decay width of the Majorana neutrino is 
$\Gamma^{tot}_{N_1}= [\Gamma(N_1 \to l \; \phi)+\Gamma(N_1 \to \bar{l} \; \phi^\dagger)]\left(1+X\right),$
with $\Gamma(N_1 \to l \; \phi)+\Gamma(N_1 \to \bar{l} \; \phi^\dagger)
= \left[\left(\lambda_\nu \lambda_\nu^\dagger\right)_{11}/ 8 \pi \right] M_1$.

In the first case, $M_1 > M_{W_R}$, the two body decay width is:
\be \Gamma_1^{2b}= \frac{g^2 \;\; M^3_1}{32 \pi M^2_{W_R}} \left(1- \frac{M^2_{W_R}}{M_1^2}\right)^2
 \left(1+ 2 \frac{M^2_{W_R}}{M_1^2}\right),\ee
which therefore implies the following dilution factor (with $a_w = \MWR^2/M_1^2$ the additional parameter which describes leptogenesis
in the present case):
\be X =\frac{g^2 v^2}{4 \mt1 M_1} \frac{\left(1-a_w\right)^2 \left(1+2 a_w \right)}{a_w}.\ee
As an estimate, taking for instance $\mt1 \sim \mathcal{O}(10^{-4}) \; eV$ and $M_1 \sim \mathcal{O} (10^{11})\; GeV$,
which in the analysis of the minimal leptogenesis model seems to be in the allowed favored range of parameters to get a 
baryonic asymmetry of the order of $n_B /s \sim \mathcal{O}(10^{-10})$;  for $a_w=1/2$, we get a dilution
factor $X \sim \mathcal{O}(10^4 \; - \;10^5)$.
Since the dilution increases for lower $a_w$, the dilution due to the two-body decay channel ruins completely the produced
asymmetry and leads to a non acceptable leptogenesis scheme.
We will not consider here the case where the gauge boson is nearly degenerate with the Majorana particle.
\medskip

For the second case, $\MWR >M_1$, the decay width of the Majorana neutrino $N_1$ in three bodies can be written as:
\be\Gamma_1^{3b}= \frac{3 g^4}{2^{10} \pi^3} \frac{M_1^5}{\MWR^4}. \label{3bd}\ee
The corresponding dilution factor, 
\be X= \frac{3 g^4 v^2}{2^7 \pi^2} \frac{1}{\mt1 M_1 a_w^2},\ee
decreases rapidly with the hierarchy and for $\mt1 \sim \mathcal{O}(10^{-4}) \; eV$ and $M_1 \sim \mathcal{O} (10^{11})\; GeV$
with $a_w \sim 10$ we only get a dilution $\sim \mathcal{O}(10)$.

We  keep here the possibility of a 
successful leptogenesis and will therefore study the dynamical evolution in this case to reveal 
the characteristic features of this realistic frame, namely additional constraints, but also new windows for the
re-heating scenario.

\section{Boltzmann equations with a minimal right-handed gauge sector}

In this section, we consider the extension of the standard Boltzmann equations for leptogenesis to the minimal
right-handed gauge sector presented in the introduction. In this model, the Majorana mass has the same origin than the 
breaking of the left-right structure through the expectation value of a scalar triplet of $SU(2)_R$. As a 
consequence, interactions mediated by the gauge sector of right-handed fermions may be relevant with respect to the
decoupling of Majorana neutrinos when decaying \cite{WRdiff}. 
The influence of those processes on the 
leptogenesis efficiency depends on the relative hierarchy between the $SU(2)_R$ gauge boson masses and the Majorana mass.

We assume here that the remaining components of the triplet scalar are heavy enough to be neglected in the evolution equations, and recall briefly the usual interactions of the scalar doublets in relation to the $(\mt1,M_1)$ parameters.

The decay through the scalar vertex $(N_1 \to l_L \phi)$ and $\Delta L=1$ diffusions mediated by the scalar doublet 
$\phi$ (Fig. \ref{phi-diffusion2}) are important 
in the Boltzmann equations with high $\mt1$. Indeed, we can easily see that (see appendix \ref{Boltz} for notations of the reduced quantities appearing in the Boltzmann equations):
\be \frac{1}{H(M_1)s}  \left( \gamma_{D, \phi} , \gamma_{\phi, s}, \gamma_{\phi, t}\right) \propto \mt1,\ee
for $\gamma_{D, \phi}=\gamma[(N_1 \to l_L \phi)+(N_1 \to \bar{l}_L \phi^\dagger)]$, 
$\gamma_{\phi, s}=\gamma[(N_1 l_L  \to \bar{t}_R Q)]$ and
$\gamma_{\phi, t}=\gamma[(N_1 t_R  \to \bar{l}_L \bar{Q})]$, the corresponding reaction densities.
They limit therefore the Majorana decoupling for growing Yukawa couplings and imply a reduction of the leptogenesis 
efficiency.

For $\Delta L=2$ diffusions (Fig. \ref{phi-diffusion1}),
 if we restrict ourselves to one Majorana neutrino in the propagator, we get:
\be \frac{\gamma_{N(, \; t)}}{H(M_1)s} \propto \mt1^2 M_1,\ee
for $\gamma_{N}=\gamma[(l_L \phi \to \bar{l}_L \phi^\dagger)]$ and 
$\gamma_{N, \; t}=\gamma[(l_L l_L \to \phi \phi^\dagger)]$.
Even though $\gamma_N$ and $\gamma_{N, \;t}$ increase with $M_1$ in this case, the mediation through heavier Majorana neutrinos
 $N_2$ and $N_3$ strengthens this behavior. The exact evaluation of this effect is actually impossible since it depends on the
$N_2$ and $N_3$ masses together with their Yukawa couplings to light neutrinos. We adopt therefore the parametrisation of 
\cite{Barbieri:1999ma, Giudice, Hambye}
to take this effect into account.

Incidentally, since our goal is mainly to highlight the possible influence of interactions mediated by the $SU(2)_R$ gauge
sector, we neglect here scatterings with light SM gauge bosons as well as thermal effects and couplings renormalization
which according to \cite{Giudice} give only $\mathcal{O}(1)$ corrections and less.

We turn now to the discussion of $SU(2)_R$ gauge bosons interactions.
As already mentioned, since  the case $M_1> \MWR$ is completely ruined by the gauge dilution of the CP asymmetry, we only
consider here interactions for $M_1 <\MWR$.
As a consequence, the dominant interactions to be added here, are 
the $W_R$ and $Z^\prime$ - mediated scatterings. The Boltzmann equations are therefore modified by the inclusion of the following terms:

\begin{minipage}[t!]{\linewidth}
\begin{center}
\includegraphics[width=.75\linewidth]{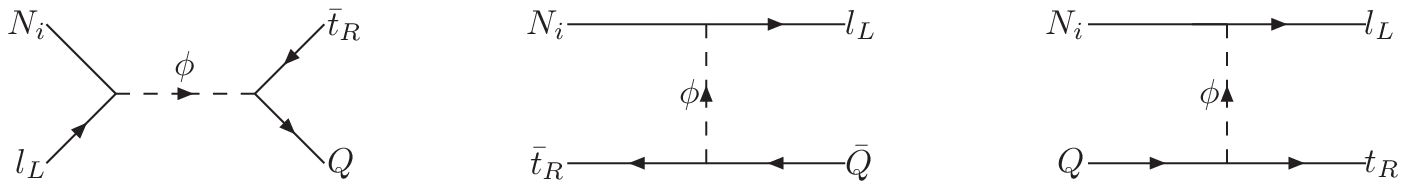}
\figcaption{$\Delta L =1$ diffusion interactions.}
\label{phi-diffusion2}
\includegraphics[width=\linewidth] {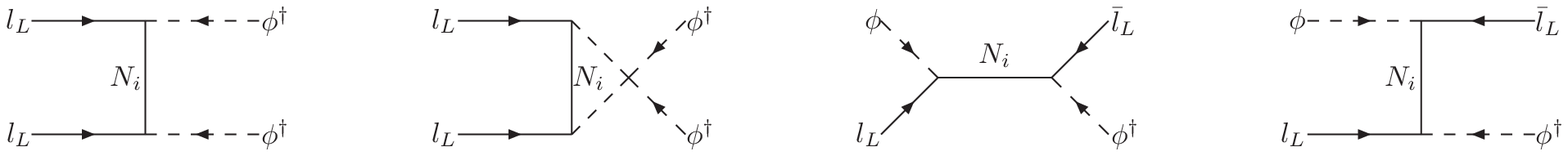}
\figcaption{$\Delta L =2$ diffusion interactions.}
\label{phi-diffusion1}
\end{center}
\end{minipage}

\medskip

\begin{minipage}[t!]{\linewidth}
\begin{center}
\includegraphics[width=.75\linewidth] {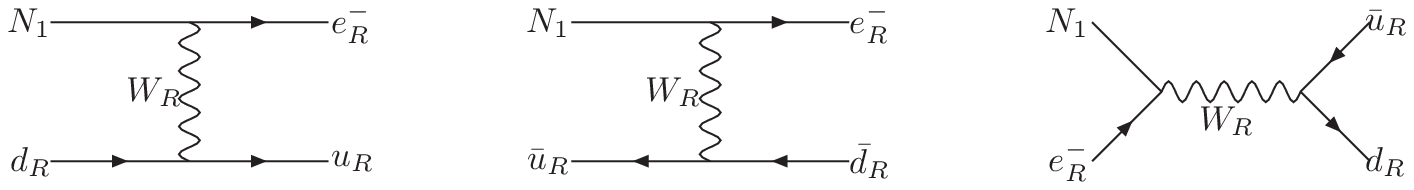}
\figcaption{diffusion interaction with one $N_1$ to be added in a gauge theory.}
\label{gauge-diffusion1}
\includegraphics[width=.75\linewidth] {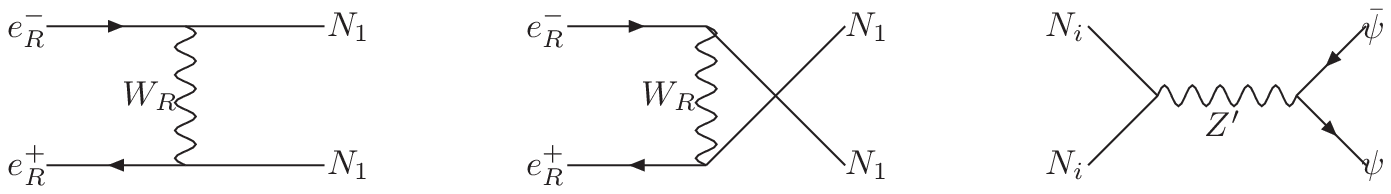}
\figcaption{diffusion interaction with two $N_1$ to be added in a gauge theory.}
\label{gauge-diffusion2}
\end{center}
\end{minipage}

\medskip

\begin{center}
\begin{figure}[t!]
\hspace{-2.5cm}
\includegraphics[width=.6 \linewidth]{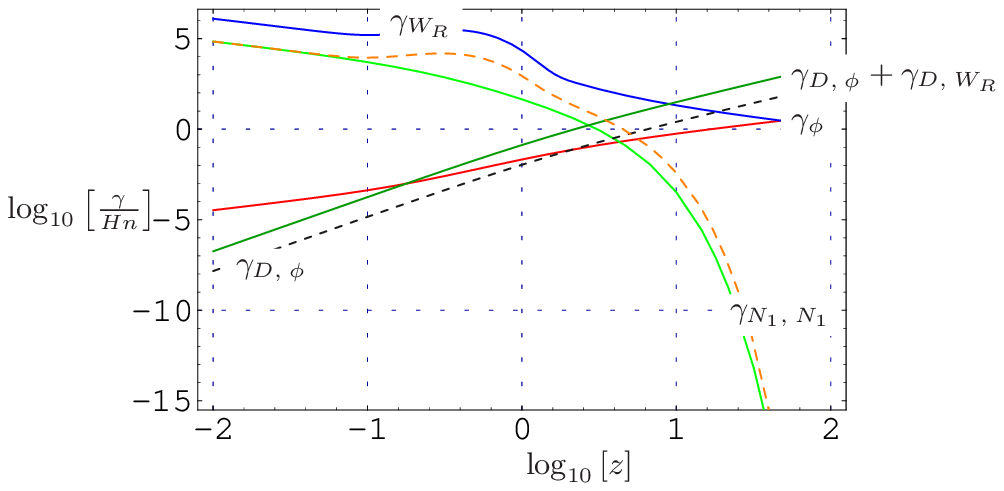}\hspace{-1.5cm}
\includegraphics[width=.6 \linewidth]{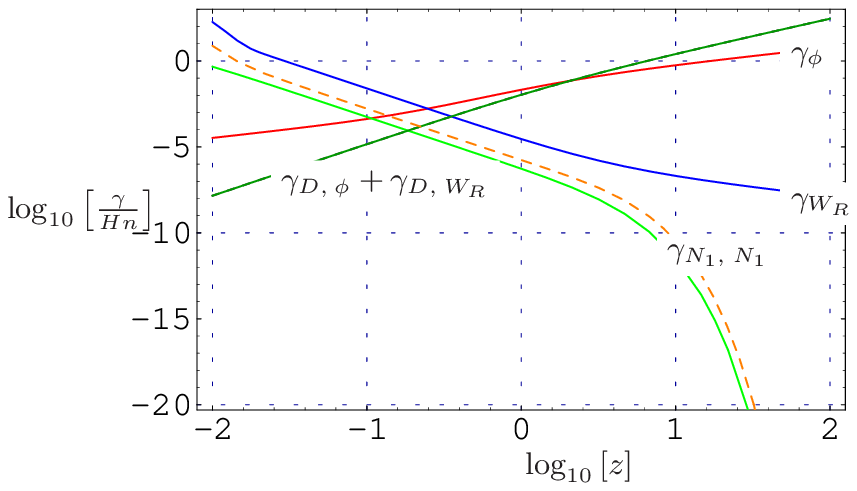}
\caption{Examples of the competition between interactions rates in the minimal case and the gauge extended case.
\textit{\underline{Left :}}
$a_w=10^2$;
 \textit{\underline{Right :}}
$a_w=10^6$, both with $\mt1= 10^{-5} eV$ and $ M_1=10^{10} \; GeV$.
The quantity $\gamma_\phi$ (resp. $\gamma_{W_R}$)
is the linear coefficient involved in the Boltzmann equation for the scatterings with an intermediate $\phi$ (resp. $W_R$),
 i.e. $\gamma_\phi = 2 \gamma_{\phi,\; s} +4 \gamma_{\phi,\; t}$
(resp. $\gamma_{W_R} =2 \gamma_{N_1,\; e_R} +2 \gamma_{N_1, \;d_R} +2 \gamma_{N_1,\; u_R}$).
The $N_1 - N_1$  scattering mediated by the $SU(2)_R$ gauge bosons is also shown; the continuous line takes
into account the $W_R$ alone while the dashed line includes also the $Z^\prime$.}\
\label{rates}
\end{figure}
\end{center}

\begin{itemize}

\item the 3-body decay channels $\gamma_{D, \; W_R}$, with the partial width given by (\ref{3bd}). Note that the out of equilibrium condition
$\Gamma^{3b} < 3 H$, yields the constraint : \be M_1 > \frac{10^{12} GeV}{a_w^2}.\label{equil3bd}\ee

\item the $\gamma_{N_1,\; e_R}= \gamma[(N_1 \;  e_R \to \bar{u}_R d_R)]$,
$\gamma_{N_1,\; u_R}= \gamma[(N_1 \;  u_R \to \bar{e}_R d_R)]$ and
$\gamma_{N_1,\; d_R}= \gamma[(N_1 \;  d_R \to e_R u_R)]$ reaction densities for all quarks families and colors, which
are added in the $Y_{N_1}(z)$ evolution equation to the standard $\Delta L=1$ diffusions (Figure \ref{gauge-diffusion1}).

\item the $\gamma_{N_1, \; N_1}=\gamma[(N_1 \; N_1 \to e_R \bar{e}_R, \;\; \psi \;\; \bar{\psi})]$ reaction density (
for all $\psi$ fermions in the SM) introducing
a quadratic term in $Y_{N_1}$ (Figure \ref{gauge-diffusion2}). This reaction however suffers of one more parameter, i.e. the $Z^\prime$ mass. We could
deal with that problem by considering the two limit cases $M_{Z^\prime}=\MWR$ and $M_{Z^\prime} \to \infty$ (shown in 
Fig. \ref{rates}). The final
leptonic asymmetry would then be in between these two cases. Nevertheless numerical computation shows only irrelevant 
differences between these cases in the final leptogenesis efficiency. We therefore present the only case
$M_{Z^\prime}=\MWR$.
One should observe moreover that scattering mediated
by the $Z^\prime$ boson can be neglected at energies far above the resonance. Actually, the
$W_R$-mediated process being in the $t$ and $u$ channels, its reduced cross-section ( $\hat{\sigma} \sim s \sigma$) logarithmically grows with the
center of mass energy while the $Z^\prime$-mediated interaction, in the $s$ channel, has a constant reduced cross section
after the resonance.

On the other hand, even though $Z^\prime$ mediated processes have already been considered in \cite{plum97}, 
the inclusion here of charged current interactions has a larger impact on the evolution of Majorana neutrinos
since their corresponding rates are leading and less suppressed for decreasing temperature.
\end{itemize}

All these processes are independent of the Yukawa couplings but are more efficient for lower $M_1$:
\be \frac{1}{H(M_1) s}
\left( \gamma_{D, \; W_R} ,\gamma_{N_1,\; e_R},  \gamma_{N_1,\; u_R},  \gamma_{N_1,\; d_R},
\gamma_{N_1, \; N_1}\right) \propto \frac{1}{M_1}. \label{gWR}\ee
They do not influence the wash-out of the leptonic number and mostly act in preventing the departure from equilibrium of
Majorana neutrinos. 

We expect a decrease of the efficiency of leptogenesis at low
$M_1$ due to these effects, depending on the ratio $a_w$, as can be expected from (\ref{equil3bd}) and (\ref{gWR}).

Finally, we write the Boltzmann equations under discussion which have to be solved numerically.
\footnote{We do not consider here an evolution equation for the leptonic number of right-handed
fermions since we assume no CP asymmetry in that sector, which is treated here at tree level.} Results are presented in the
next section.
\begin{eqnarray}
\frac{s H(M_1)}{z}\frac{dY_{N_1}}{dz} &=&- \left[\gamma_{D,\; \phi}+\gamma_{D,\; W_R}
+2 \gamma_{\phi,\; s} +4 \gamma_{\phi,\; t} +2 \gamma_{N_1,\; e_R} +2 \gamma_{N_1, \;d_R} +2 \gamma_{N_1,\; u_R} \right]
 \left(\frac{Y_{N_1}}{Y_{N_1}^{eq}}- 1\right) - \gamma_{N_1,\; N_1}  \left[\left(\frac{Y_{N_1}}{Y_{N_1}^{eq}}\right)^2- 1\right]\\
\frac{s H(M_1)}{z}\frac{dY_{L}}{dz} &=& - \epsilon_1  \gamma_{D,\; \phi} \left[\frac{Y_{N_1}}{Y_{N_1}^{eq}}-1\right]
- \frac{Y_L}{Y_L^{eq}} \left[\frac{\gamma_{D,\; \phi}}{2} +2 \gamma_{N}+ 2 \gamma_{N, \; t}
+2 \gamma_{\phi,\; t} + \frac{Y_{N_1}}{Y_{N_1}^{eq}} \gamma_{\phi,\; s}\right]
\end{eqnarray}

\section{Matter asymmetry and leptogenesis efficiency}\label{Matter}

The present baryonic asymmetry of the universe is quantified by the ratio between the net number of baryons and the number
of photons in the universe. From nucleosynthesis analysis and measurements on the cosmic microwave background, one can
deduce the very smallness of this asymmetry, i.e. according to the recent WMAP results \cite{WMAP}:
\be \frac{n_B}{n_\gamma} = \left(6.1 ^{+0.3}_{-0.2} \right) 10^{-10}.  \label{WMAP}\ee

In order to explain this quantity through the leptogenesis process, once a lepton asymmetry is produced, we
require the convertion of this non-zero lepton number in a non-zero baryon number \cite{Kuzmin:1985mm}. This convertion is provided by
non-perturbative electroweak interactions, as exemplified by sphalerons  which take place around the electroweak phase transition and conserve $B-L$ 
while  violating both $B$ and $L$. A complete conversion would lead to $B_{final}=-1/2 L_{initial}$ but is
usually evaluated to $B_{final}=-28/79 L_{initial}$ \cite{Shapo}.

Now, in the most favorable case, the maximal lepton asymmetry produced by the Majorana neutrino decay can be estimated
as $Y_L^{ final}= \epsilon_1 Y_{N_1}^{eq}(init.)$, where $Y_{N_1}^{eq}(init.)$ is the initial equilibrium $N_1$ abundance. 
However, the leptogenesis efficiency is not always maximal and depends mainly on
the importance of the departure from equilibrium and the possible wash-out of 
the produced lepton asymmetry by $\Delta L \neq 0$ diffusions. We can easily
write an analytical expression for the efficiency using the integral form of the
Boltzmann equation for the lepton number: \be \eta_{
eff}=\int_0^\infty dt \frac{z
(Y_{N_1}-Y_{N_1}^{eq})}{s H(M_1) Y_{N_1}^{eq}(init.)^2} \gamma_{D,\; \phi}\;
\mbox{Exp}\left[-\int_t^\infty  dx \;\; W_L(x)\right],\ee 
for 
$W_L(z)=\frac{z}{s H(M_1) Y_{N_1}^{eq}} \left(\frac{\gamma_{D,\; \phi}}{2}
+2 \gamma_{N}+ 2 \gamma_{N, \; t} +2 \gamma_{\phi,\; t} +
\frac{Y_{N1}}{Y_{N1}^{eq}} \gamma_{\phi,\; s}\right)$ the wash-out terms of the leptonic asymmetry. 
The efficiency does not depend on
the CP asymmetry and characterizes therefore the intrinsic dynamics of the
leptogenesis process in a given theory.

As a consequence, depending on the parameters, we can express the final leptonic asymmetry including the efficiency by:
\be\frac{n_L^{final}}{s}=Y_L^{final}= \epsilon_1 Y_{N_1}^{eq}(init.) \eta_{eff}.\ee
Finally, the baryons to photons ratio is evaluated in term of leptogenesis quantities to :
\be\frac{n_B}{n_\gamma} \simeq 7 \frac{28}{79} Y_{B-L}= -\frac{196}{79} \epsilon_1 Y_{N_1}^{eq}(init.) \eta_{eff},
\label{nb}\ee
where the factor $7$ holds for $s/n_{\gamma}$.

\medskip

We present results for the leptogenesis efficiency $\eta_{eff}$ as a function of 
$\mt1$, $M_1$ and $a_w$ the $W_R$ to Majorana mass squared ratio in Figure \ref{eta}.

 Both thermal and zero initial Majorana abundance scenarios are
shown but, however, with respect to the minimal leptogenesis scheme which is significantly more constrained by a zero initial Majorana
abundance, the extended gauge model shows less dependency according to these cases even for a sizable hierarchy. Indeed, as one can easily
expect, the creation of heavy Majorana neutrinos is not anymore a problem once they are allowed to take part in gauge interactions. This effect remains up to ratios of the order $a_w \propto 10^6$ in the relevant $M_1$ range.

The second interresting consequence of the inclusion of $W_R$ mediated diffusions is a drastic 
reduction of leptogenesis efficiency for low Majorana masses which increases dramatically for low $a_w$-ratio.
This comes from both delayed (or impeded) decoupling of Majorana neutrinos through
additional diffusions and decay channels: the final leptonic asymmetry is therefore reduced.

\medskip

Now, in order to discuss the baryon number, we have to link observables on neutrino oscillations and the CP asymmetry in
the decay of the lightest Majorana (\ref{eH}). As already proposed by \cite{Giudice} and for the purpose of comparison,
we consider the sample case where $m_3=\mbox{max}(\mt1, \sqrt{\Delta
m^2_{atm}})$ with $m_3^2-m_1^2=\Delta m^2_{atm}$. 
Combining equations (\ref{eH}) and (\ref{nb}) therefore provides an upper estimate of the baryons to photons
ratio. 

Figure \ref{Masymtot} shows the iso-$n_b/n_\gamma$ curves in the $(\mt1, M_1)$ plane obtained for
different $a_w$-ratios, assuming the central value of baryon asymmetry obtained from WMAP (\ref{WMAP}). Successful leptogenesis occurs in the region bounded by the
curves.

We can then extract a lower limit on the Majorana mass according to the $W_R$ to Majorana mass squared ratio:

\begin{center}
\begin{tabular}{c|c}
$a_w= \MWR^2/M_1^2$ & $M_1>$ (GeV)\\
\hline $10^4$ & $\sim  10^9 $ \\
 $10^2$ & $\sim  10^{10} $ \\
 $10$ & $\sim   10^{11} $ \\
  $2$ & $\sim  10^{12} $ \\
\end{tabular}
\end{center}
As already mentioned above, the case with zero initial Majorana abundance scenario (after assumed re-heating) is almost indistinguishable in our framework from the case of a thermal initial
abundance. Consequently, in the case of re-heating, we allow, despite the dilution, for broader windows in some domains of the parameters .

For comparison, the minimal (non-gauged Majorana) model curve for an initial thermal Majorana abundance co{\"\i}ncides in practice with our curve for $a_w = 10^6$ (the latter, irrespective of the thermal or re-heating scenario).

\section{Conclusion}

Inspired by a grand-unified framework necessary to understand the presence of heavy Majorana particles,
 we considered the addition of a minimal gauge sector for right-handed neutrinos and derived its consequences 
on baryon number
in both the thermal decay and re-heating scenarios.

The results are :
\begin{itemize}
\item  a diluted CP asymmetry due to new Majorana decay channels through gauge bosons, 
restricting successful leptogenesis to the case $\MWR > M_1$ \cite{ling-frere};

\item a reduction of the leptogenesis efficiency for low $W_R$ to Majorana mass squared ratio
( for $a_w  < 10^4$);

\item in the re-heating scenario, an enhanced production of Majorana neutrinos thanks to their gauge interactions 
( up to $a_w \sim 10^6$).
\end{itemize}

The consequences from the baryon to photon ratio on the neutrino mass parameters with respect to the minimal scheme
is a higher Majorana mass for decreasing $W_R$ to Majorana mass ratio and a broader window in the case of zero 
initial Majorana abundance (see Figure \ref{Masymtot}).

\section*{Acknowledgments}
I would like to thank especially J.-M. Fr\`ere for his advice and support all along this work.
I also thank P. Aliani, M. Fairbairn, F.-S. Ling and M.H.G. Tytgat for interesting discussions.
This work is supported in part by IISN, la Communaut\'{e} Fran\c{c}aise de Belgique (ARC),
 and the belgian federal government (IUAP-V/27).

\appendix

\section{Boltzmann equations}
\label{Boltz}

We quickly summarize the conventions we used in the text with respect to the evolution equations of leptogenesis.
The particle density is define as :
\be n_a = g_a \int \frac{d^3 p}{(2 \pi)^3} f_a(\bf{p}),\ee
where $g_a$ is the number of internal degrees of freedom of the particle and $f_a$ is the corresponding statistics.
We assume here the Maxwell-Boltzmann statistics, so that we can re-write:
\be n_a =\frac{g_a M_a^2 T}{2 \pi^2} K_2(M_a/T), \ee
which is the equilibrium number density ($K_n(x)$ is the modified Bessel function of order $n$).

The evolution equations can then be written in term of the number of particles per comoving volume, $Y_a(z)=n_a(z)/s(z)$ (where $s(z)$ is the
entropy density, $z= M_1/T$), which is not affected by the expansion of the universe :
\be \frac{dY_a}{dz}=- \frac{z}{H(z=1) s(z)} \sum_{a,I,J} \left[\frac{Y_a Y_I}{Y_a^{eq} Y^{eq}_I} \gamma(a I \to J)- \frac{Y_J}{Y_J^{eq}} \gamma(J \to a I)\right].\ee
They constitute in general a set of coupled differential equations for all particles species in the model.
The $\gamma_{(a I \to J)}$ are the reaction densities for the $(a I \to J)$ interactions which is given for a decay process by:
\be \gamma_D= n_a^{eq} \frac{K_1(z)}{K_2(z)} \Gamma_D,\ee
(where $\Gamma_D$ is the usual decay width) and for a two-body scattering by ($x=s/M_1^2$, $\sqrt{s}$
the center of mass energy):
\be \gamma(a I \to J)= \frac{M_1^4}{64 \pi^4 z}
\int_{x_{min}}^\infty dx \; \hat{\sigma}(a I \to J)(x) \sqrt{x} K_1(z \sqrt{x}).\ee

\section{Reduced cross sections}
The reduced cross sections are the kinematic ingredients from Feynman graphs that are involved in Boltzmann equations.
For a given process, they can be computed through:
\be \frac{d\hat{\sigma}}{dt} = \frac{1}{8 \pi s} | \mathcal{M}(a I \to J)|^2,\ee
in terms of the Mandelstam variables ( $|\mathcal{M}(a I \to J)|^2$ being the squared amplitude summed over all
internal degrees of freedom of $a$, $I$ and $J$). The reduced cross section is related to the usual cross section
by $\hat{\sigma}(s) =(8/s) [(p_a . p_I)^2 -m^2_a m^2_I] \sigma(s)$.

We list below the reduced cross section used for the relevant minimal model interactions 
(see e.g. \cite{Luty:un},  \cite{plum97}, \cite{Giudice}, \cite{Pilaftsis:2003gt}):
\be x=\frac{s}{M_1^2}, \qquad a_\Gamma = \frac{\Gamma_{N_1}^{D \; 2}}{M_1^2}, \qquad D_{N_1}= \frac{1}{x-1+i \sqrt{a_{\Gamma}}},
 \qquad D_{N_1}^{2 \;\; subs}= \frac{(x-1)^2-a_{\Gamma}}{\left[(x-1)^2+a_{\Gamma}\right]^2},\ee
where $D_{N_1}^{2 \;\; subs}$ is the $N_1$ propagator amplitude without its resonant part as suggested in \cite{Giudice} which
avoid the double counting of real Majorana's in the inverse decay and in the $\Delta L=2$ diffusions.

\underline{$\Delta L =2$ (mediated by the $N_1$ alone):}

\begin{itemize}

\item $l_L \;\; \phi \to \bar{l}_L \;\;\phi^\dagger$
\be \hat{\sigma}_{N_1}=\frac{1}{4 \pi} \left(\lambda \lambda^\dagger\right)^2_{11} \left[ x \; D_{N_1}^{2 \;\; subs}
+2 \left(1-\frac{1}{x} \log[x+1]\right)
+ 4 \mathcal{R}e\left[D_{N_1}\right]\left(1-\frac{x-1}{x} \log[x+1]\right)\right]\ee

\item $l_L \;\; l_L \to \phi^\dagger \;\; \phi^\dagger$
\be \hat{\sigma}_{N_1}=\frac{1}{2 \pi}
\left(\lambda \lambda^\dagger\right)^2_{11} \left[ \frac{x}{(x+1)}+ \frac{2}{2+x} \log[x+1]\right] \ee

\end{itemize}

\underline{$\Delta L=1$:}
\begin{itemize}

\item $N_1 \; \; l_L \to \bar{t}_R \; \; Q$
\be \hat{\sigma}= \frac{3}{4 \pi} \left(\lambda \lambda^\dagger\right)_{11} \frac{m_t^2}{v^2} \left( \frac{x-1}{x}\right)^2 \ee

\item $ N_1 \; \; t_R \to  \bar{l}_L \; \; Q$
\be \hat{\sigma}= \frac{3}{4 \pi} \left(\lambda \lambda^\dagger\right)_{11} \frac{m_t^2}{v^2} 
\left( \frac{x-1}{x} + \frac{1}{x} \log[\frac{x-1+a_{\phi}}{a_{\phi}}] \right),\ee
\end{itemize}where $a_{\phi}= m^2_\phi/M_1^2$ is an infra-red regulator chosen such that $m_\phi \sim 800 GeV$ \cite{plum97}.

Since we include extra gauge bosons in the model, we list below the reduced cross-sections for the relevant $SU(2)_R$ diffusions:
\be a_{w( \; Z^\prime)}=\frac{M^2_{W_R( \; Z^\prime)}}{M_1^2} \qquad a_{\Gamma_{W_R( \; Z^\prime)}}=\frac{\Gamma^2_{W_R( \; Z^\prime)}}{M_1^2}.\ee
\begin{itemize}
\item $N_1 \; \; d_R \to e_R \;\; u_R$
\be \hat{\sigma}= \frac{9 g^4}{8 \pi a_w} \frac{\left( x-1\right)^2}{\left(x+a_w-1\right)},\ee

\item $N_1 \;\; u_R \to \bar{e}_R \;\; d_R$
\be \hat{\sigma}= \frac{9 g^4}{8 \pi x} \int^0_{(1-x)} \!\!dt \;\; \frac{(x+t-1) (x+t)}{(t-a_w)^2}, \ee

\item $N_1 \;\; e_R \to \bar{u}_R \;\; d_R$
\be \hat{\sigma}=
\frac{9 g^4}{8 \pi x \left[ \left(x-a_w\right)^2+ a_w \; a_{\Gamma_{W_R}}\right]} \int^0_{(1-x)} \!\!dt \;\;(x+t) (x+t-1),\ee

\item $N_1 \;\; N_1 \to e_R \; \; \bar{e}_R, \;\; \psi \;\; \bar{\psi}$
\be \hat{\sigma}_{W_R}=\frac{g^4}{8 \pi x} \int_{t_{0}}^{t_{1}}dt \left[ \frac{(x+t-1)^2}{\left[t-a_w\right]^2}
+\frac{(1-t)^2}{\left[2-a_w-x-t\right]^2} - \frac{2 x}{\left[t-a_w\right] \left[ 2-a_w-x-t \right]}\right],\ee

\be
\hat{\sigma}_{Z^\prime}=\frac{g^4 R^2_{N_1}}{3\;\; 2^5 \pi c^4_w} \frac{\sqrt{x \left( x-4\right)} \left(x-1\right)^2}
{\left[\left(x-a_{Z^\prime}\right)^2 + a_{Z^\prime} a_{\Gamma_{Z^\prime}}\right]} \left[ 9 R^2_{u_R}+ 9 R^2_{d_R} +3 R^2_{e_R}
+6 L^2_{e_L}+ 18 L^2_{q_L} \right]
\ee
\be
R_f = \sqrt{c^2_w - s^2_w} \left( 2 T_R^3 - s^2_w \frac{(B-L)}{c^2_w -s^2_w}\right), \qquad
L_f= - \frac{s^2_w (B-L)}{ \sqrt{c^2_w -s^2_w}}
\ee
\end{itemize} with $t_0(t_1)$ the kinematic limits on the $t$ variable. The minus sign in the interference term of 
$\hat{\sigma}_{W_R}$ is due to the relative sign between  the $t$ and the $u$ diagrams coming from Fermi-Dirac statistic.
Note that for $N_1 \;\; N_1 \to e_R \; \; \bar{e}_R$ interferences between $s$ channel with an intermediate $Z^\prime$ and 
$t$ and $u$ channels mediated by a $W_R$ cancel out for the same reason.

We also write the $SU(2)_R$ gauge vertices added to the minimal model of leptogenesis ( with
$\Psi^{\left(R, L\right)} 
=(\begin{array}{cc} \psi^{\left(R, L\right)}_u & \psi^{\left(R, L\right)}_d \end{array})^T$) : 

\be \frac{g}{\sqrt{2}} \left(\bar{\psi}^R_u W_R^- \!\!\!\!\!\!\!\!\!\!\!\!\!/   \;\;\;\;\;\psi^R_d +h.c. \right),\ee

\be
g \frac{\sqrt{c_w^2-s_w^2}}{c_w} \bar{\Psi}_R \;\;
Z^{\prime} \!\!\!\!\!\!\!\!/  \;\;\left[T_R^3 - \frac{B-L}{2} \frac{s^2_w}{c^2_w -s^2_w} \right] 
\Psi_R + \Psi_R \to \Psi_L,
\ee
where $s_w=\sin{\theta_w}=- \frac{g^\prime}{\sqrt{g^2 +2 g^{\prime 2}}}$, $W^\pm_R = \frac{W_R^1 \mp i W_R^2}{\sqrt{2}}$,
 $Z^\prime = \frac{1}{c_w} \left[ \sqrt{c_{2w}} \;  W^3_R + s_w \; B\right]$, $g$ and $g^\prime$ respectively 
the $SU(2)$'s and $U(1)$ couplings, $W^a_R$ the a-th gauge component of $SU(2)_R$ and $B$ the $U(1)$ gauge field.


\begin{figure}[h!]
\vspace{-1.5cm}\begin{center}
\begin{narrow}{-1cm}{1cm}

\hspace{-.1 \linewidth}\includegraphics[width=.6\linewidth]{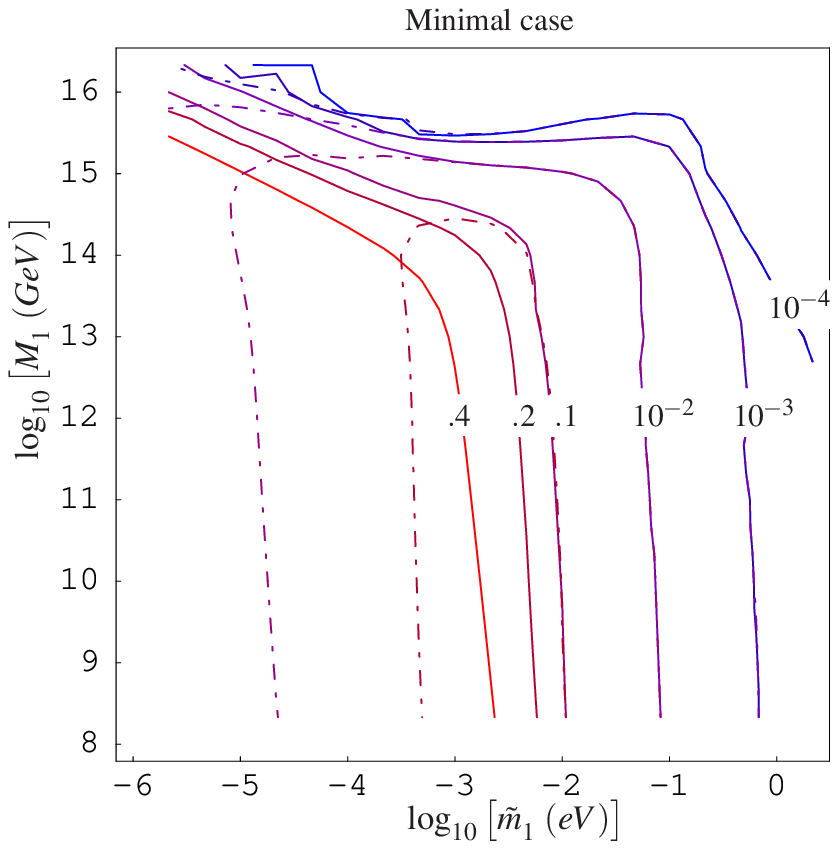}
\hspace{-.1 \linewidth}\includegraphics[width=.6 \linewidth]{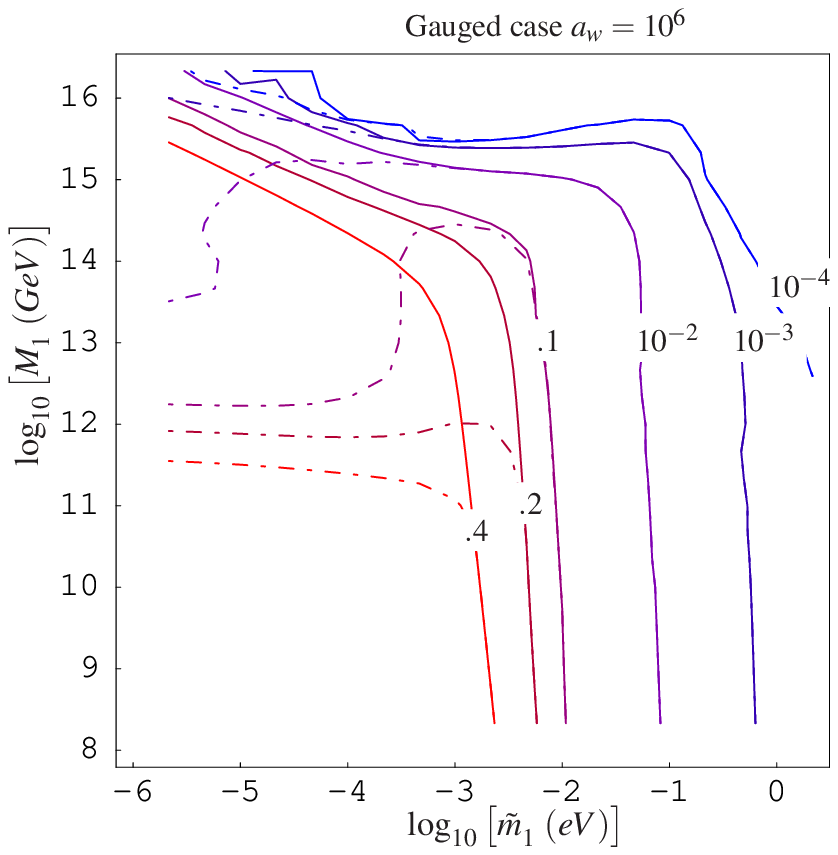}

\hspace{-.1 \linewidth}\includegraphics[width=.6 \linewidth]{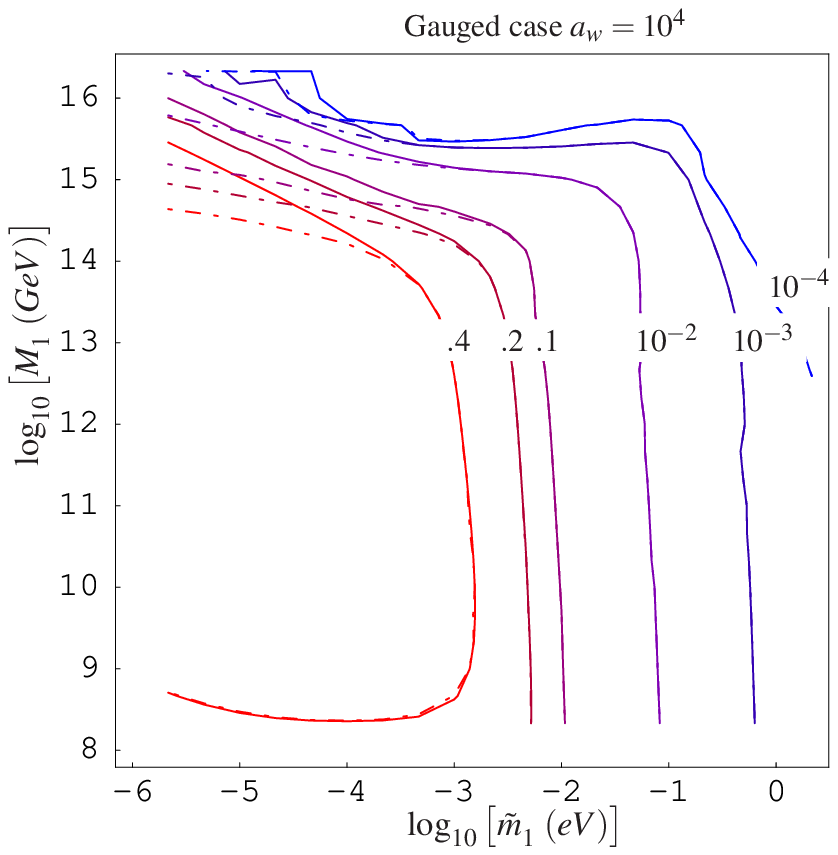}
\hspace{-.1 \linewidth}\includegraphics[width=.6 \linewidth]{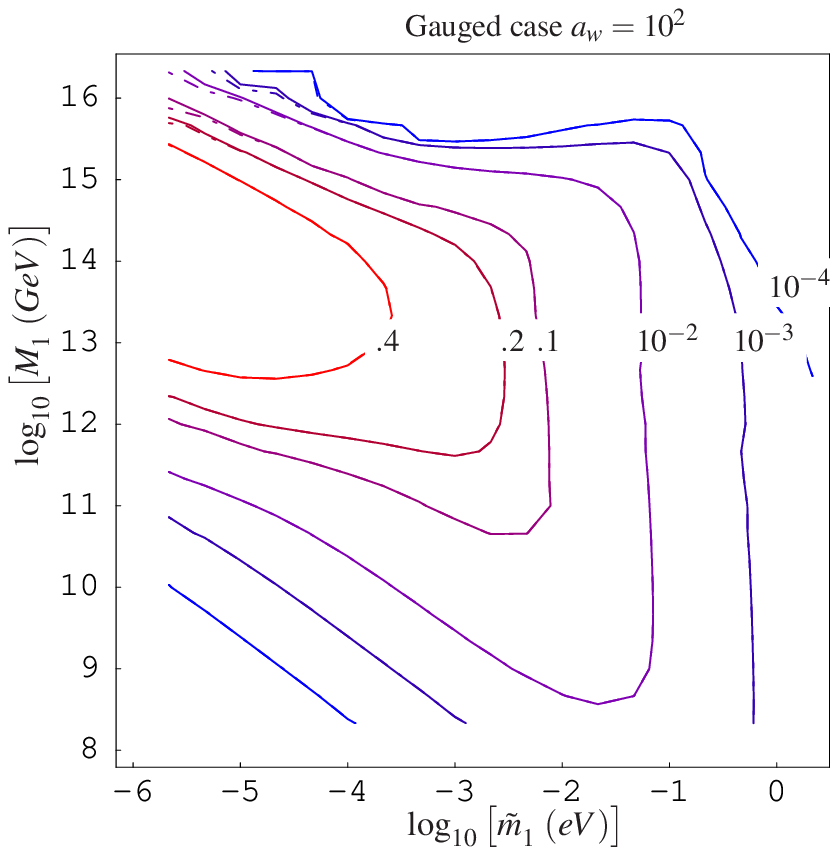}

\hspace{-.1 \linewidth}\includegraphics[width=.6 \linewidth]{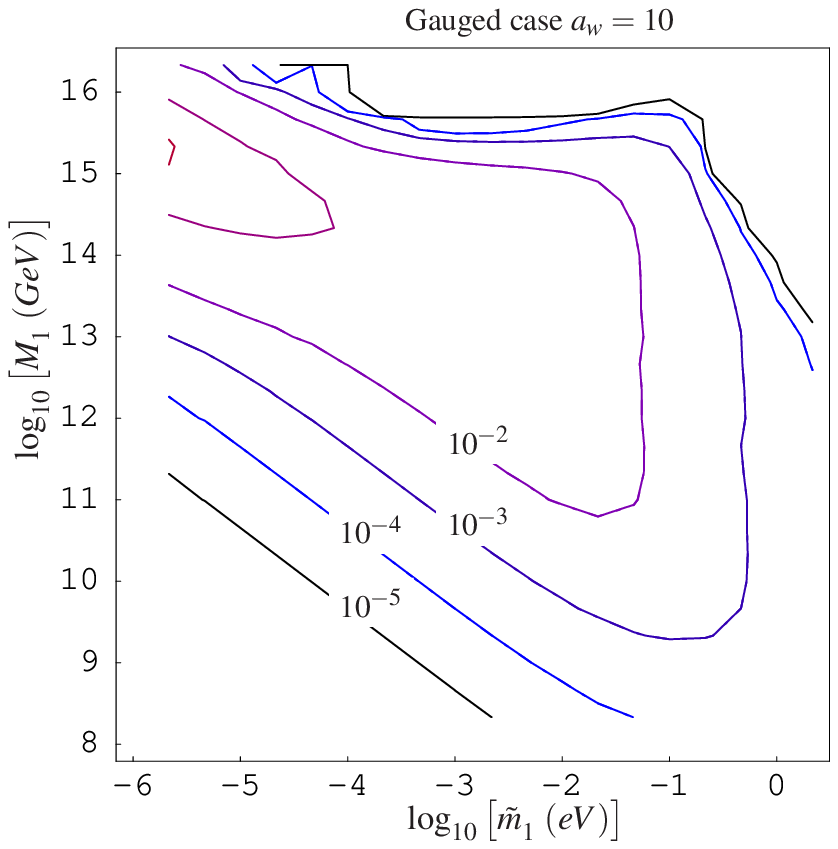}
\hspace{-.1 \linewidth}\includegraphics[width=.6 \linewidth]{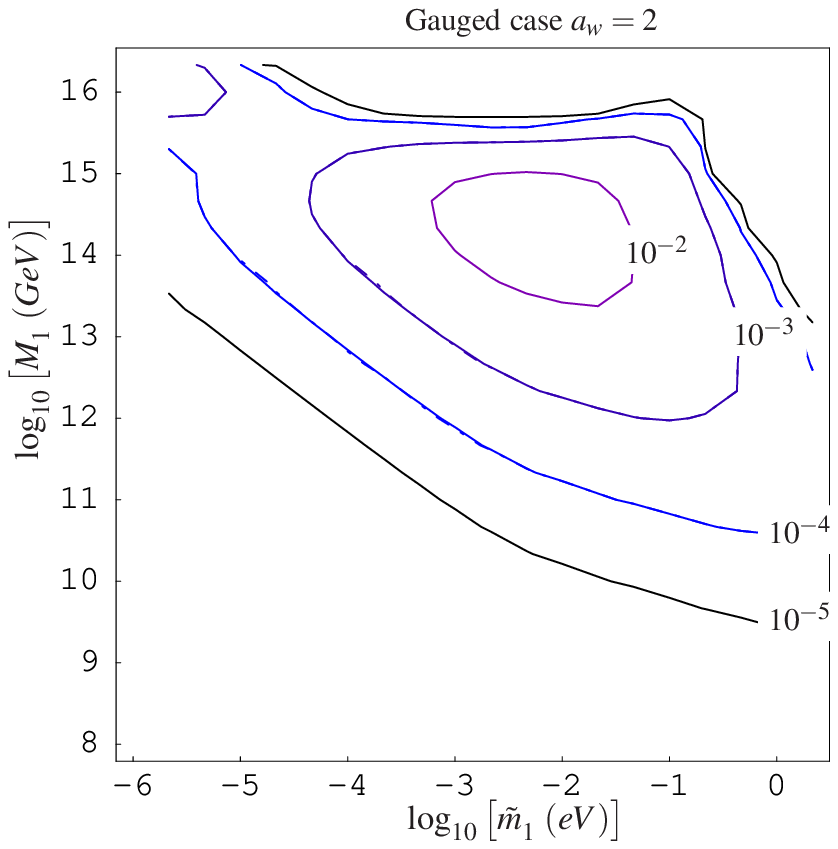}
\end{narrow}
\caption{Efficiency of leptogenesis for the minimal case $a_w \to \infty$ and
the extended case to a minimal right-handed gauge sector for $
a_w=\MWR^2/M_1^2 =10^6, \; 10^4, \; 10^2, \;10, \;2$.
The continuous line is the efficiency for a thermal initial Majorana abundance while the
dashed-doted line is for a zero initial Majorana abundance.}
\label{eta}
\end{center}
\end{figure}

\begin{figure}[t!]\begin{center}
\hspace{-5cm}
\includegraphics[width=\linewidth]{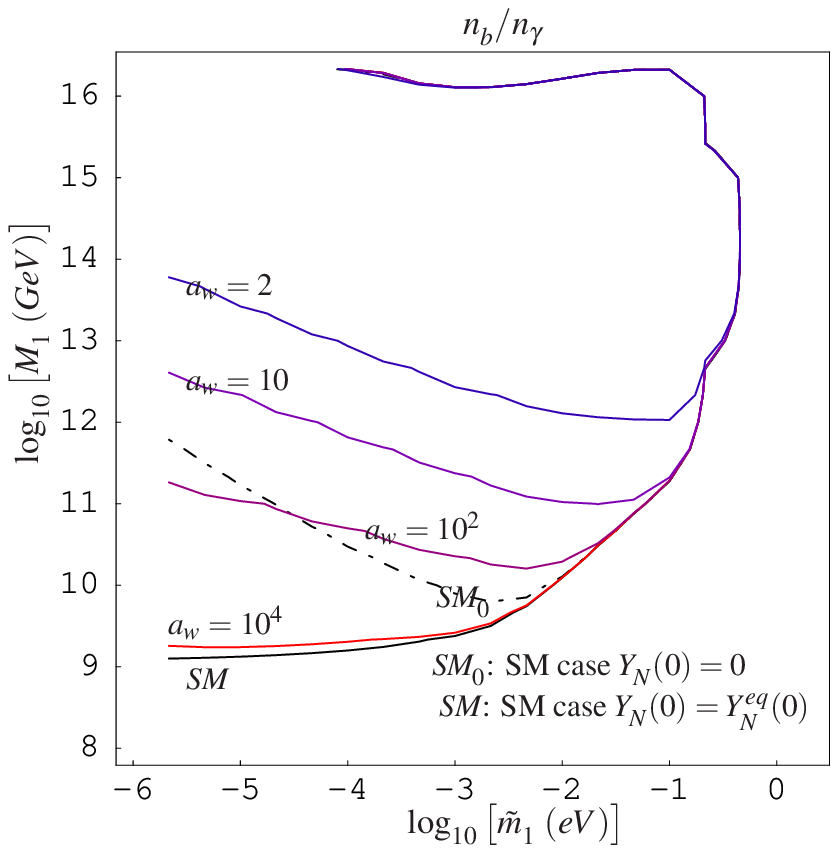}
\caption{Limits on the baryon to photon ratio in an extended model 
to a minimal right-handed gauge sector with
$a_w=\MWR^2/M_1^2 =10^4, \;\; 10^2, \;\;10, \;\; 2$,  from bottom to top (red to blue).
The minimal SM case is also shown in black:
the continuous line is for a thermal initial Majorana abundance while the
dashed-doted line is for a zero initial Majorana abundance.
The matter asymmetry is higher than the WMAP value in the region bounded by the curves.}
\label{Masymtot}
\end{center}\end{figure}

\end{document}